\documentstyle[twocolumn,aps,epsf]{revtex}

\tightenlines

\begin{document}


 \wideabs{
\title{Large scale numerical investigation of excited states in 
poly(phenylene)}

\author{Robert J. Bursill$^{1\ast}$ and William Barford$^{2\ast\ast}$}

\address{ $^1$School of Physics, The University of New South Wales, Sydney, 
NSW 2052, Australia. \\
$^2$Department of Physics and Astronomy, The University of Sheffield, \\ 
Sheffield, S3 7RH, United Kingdom. }

\maketitle

\begin{abstract}

A density matrix renormalisation group scheme is developed, allowing for the 
first time essentially exact numerical solutions for the important excited 
states of a realistic semi-empirical model for oligo-phenylenes. By 
monitoring the evolution of the energies with chain length and comparing 
them to the experimental absorption peaks of oligomers and thin films, we 
assign the four characteristic absorption peaks of phenyl-based polymers. 
We also determine the position and nature of the nonlinear optical states 
in this model.

\end{abstract}


} \narrowtext

The phenyl-based conjugated polymers have attracted immense interest from 
physicists and chemists as a result of the discovery of electroluminescence 
in poly(phenylenevinylene) (PPV) \cite{electroluminescence}. In particular, 
a great deal of theoretical and experimental effort has been devoted to 
understanding the important neutral excitations driving nonlinear optical 
processes, the positions of triplet states and the nature 
of exciton binding and decay 
\cite{binding_debate,chandross1,chandross2,rice,weibel}. Correlation 
between $\pi$-electrons has proved to be important in obtaining an accurate 
description of excited states in conjugated systems. It is thus desirable 
to have accurate numerical solutions of correlated electron models for 
these systems (without recourse to uncontrolled approximation schemes such 
as configuration interaction approximation schemes, or perturbative schemes 
starting from a noninteracting molecular orbital or $k$-space band 
description). In this paper we provide, for the first time, accurate 
results for a realistic semi-empirical model for poly(phenylene) and its 
oligomers, a model system for phenyl-based conjugated polymers.

\begin{figure}[htbp]
\centerline{\epsfxsize=7.5cm \epsfbox{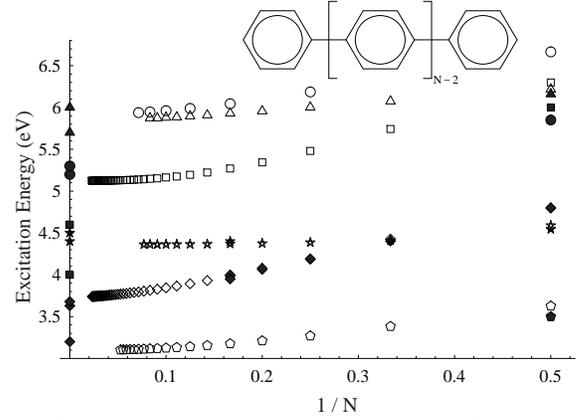}}
\caption{
The transition energies as a function of inverse chain length: 
$1^3B_{1u}^+$ (pentagons), $1^1B_{1u}^-$ (diamonds), $1^1B_{2u}^+$ (stars), 
$2^1A_g^+$ (squares), $1^1B_{2u}^-$ (circles) and the localised intra-phenyl $^1B_{1u}^-$ (Frenkel) state (triangles). The filled symbols are the 
experimental values for biphenyl (see ref.\ \protect\cite{bursill98}), 
oligomers ($N = 3,\ldots,6$ \protect\cite{oligomers} and $N = 6$ 
\protect\cite{zojer}) and thin film polymers ($N = \infty$) 
\protect\cite{oligomers,absorption,lane}. The inset shows the 
oligo-phenylene geometry.
}
\label{energies1}
\end{figure}

As our starting point for describing poly(phenylene), we adopt the Pariser-Parr-Pople (P-P-P) model \cite{pariser},
\begin{eqnarray}
{\cal H}
& = &
- \sum_{<ij>\,\sigma} t_{ij}
	\left[ c_{i\sigma}^{\dagger} c_{j\sigma} + \rm{ h.c.} \right]
\nonumber
\\
& &
+\; U \sum_{i} 
	\left(n_{i\downarrow}-1/2\right)
	\left(n_{i\downarrow}-1/2\right)
\nonumber
\\
& &
+\; \frac{1}{2} \sum_{i\neq j} V_{ij} (n_i - 1)(n_j - 1),
\label{hamiltonian}
\end{eqnarray}
where $<>$ represents nearest neighbors, $c_{i\sigma}$ destroys a $\pi$-electron on conjugated carbon atom $i$, $n_{i\sigma} = 
c_{i\sigma}^{\dagger} c_{i \sigma}$ and $n_i = n_{i\uparrow} + 
n_{i\downarrow}$. We use the Ohno parameterisation for the Coulomb 
interaction,
$ V_{ij} = U / \sqrt{ 1 + (U r_{ij}/14.397)^2 }$, where $r_{ij}$ is the 
inter-atomic distance in \AA ~and $U = 10.06$ eV \cite{bursill98}. The 
transfer integrals ($t_{ij}$) and bond lengths are $2.539$ eV and $1.4$ \AA 
~respectively for phenyl bonds, and $2.22$ eV and $1.51$ \AA ~respectively 
for single bonds \cite{bursill98}. In the following we consider oligophenyls
of $N$ repeat units (phenyl rings), as depicted in Fig.\ \ref{energies1}.

When solved accurately, the P-P-P theory has proved remarkably accurate in 
describing many of the low-lying excitations in a wide range of conjugated 
molecules \cite{bursill98,soos}. Given such debates as to the nature of 
exciton binding in PPV and the related issues of interpreting various 
spectral features \cite{binding_debate,chandross1,chandross2,rice,weibel}, 
it is intriguing as to what the P-P-P theory predicts for large phenyl 
systems. Exact diagonalisation is currently restricted to systems with two 
phenyl rings. The advent of the density matrix renormalisation group (DMRG) 
method \cite{dmrg} has been fortuitous in that it has the potential to 
provide effectively exact numerical solutions of the P-P-P theory for 
systems with hundreds of conjugated carbon atoms. The DMRG has obstensively 
been restricted to systems with acetylene-like (simple, linear chain) 
geometry \cite{pa,barford00} to date. This is because it has not been 
possible to keep a complete basis for the phenyl repeat unit (4096 basis 
states) as well as sufficient system and environment block states. In this 
work we develop an accurate method for treating the phenyl block based on 
local Hilbert space optimisation \cite{barford00,zhang98}.

Our approach starts with the assumption that an optimised local Hilbert 
space prepared in a small system should also be suitable for a large 
system. We employ the following procedure:
\begin{enumerate}

\item
The P-P-P model is solved exactly for the 12-site ($N = 2$) biphenyl 
molecule. The right-hand ring acts as the environment for the left-hand 
ring. A reduced density matrix for the left-hand ring is constructed by 
performing a partial trace of the target state(s) projection operator over 
the states of the right-hand ring. The $m_A$ optimal states for the left-hand ring (i.e.\ the optimised local Hilbert space) are those density 
matrix eigenstates with the highest eigenvalues. These are the natural many 
body states of the phenyl ring at the end of an oligo-phenylene molecule, in 
analogy to the natural one-electron orbitals of quantum chemistry.

\item
For phenyl rings added to the middle of the lattice during the DMRG 
iterations (i.e.\ rings with neighbours to the left and right), we apply 
periodic boundary conditions to the biphenyl molecule to again construct an 
optimal basis, of which the $m_B$ most important states are retained. We 
note that this method of determining a suitable repeat unit Hilbert space 
from a small, exactly diagonalisable system is employed at the first stage 
of the ``four-block'' DMRG method \cite{4block}.

\item
Once the local Hilbert space has been optimised the usual inifinite lattice 
DMRG approach is used to build up oligophenyls. We use a three-block 
infinite lattice method, the environment block being the reflected system 
block. This enables us to apply the full symmetries of the Hamiltonian, 
which are the two planes of reflection, and particle-hole and spin-flip 
symmetries. We use the symbols $A_g$ and $B_{1u}$ to denote states that are 
symmetric under long-axis reflection, the $A_g$ states being symmetric 
under short axis reflection, the $B_{1u}$ states being antisymmetric under 
short axis reflection (polarised along the molecular axis). We use the 
symbol $B_{2u}$ to denote states that are polarised perpendicular to the 
molecular axis, i.e., states which are symmetric under long axis reflection 
and antisymmetric under short axis reflection. The spin-flip symmetry is 
used to distinguish singlet and triplet states, i.e., singlet states in the 
$A_g$ sector are denoted $^1A_g$ whilst triplet states are denoted $^3A_g$. 
Finally, the particle-hole symmetry of states is denoted by a ``$+$'' or 
``$-$'' superscript. For example, the second lowest state in the $A_g$ 
singlet sector with positive particle-hole symmetry is denoted $2^1A_g^+$.

\item
Oligophenyls with an even number of phenyl rings are constructed by 
performing DMRG iterations with no repeat unit (i.e., just with a system 
and environment block).

\end{enumerate}

\begin{figure}[htbp]
\centerline{\epsfxsize=7.5cm \epsfbox{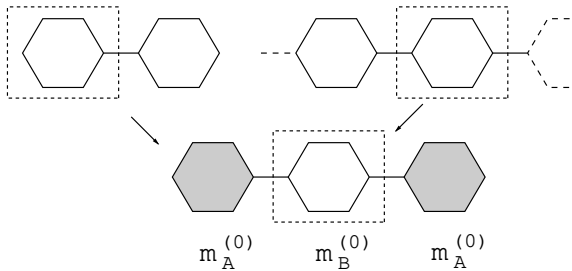}}
\caption{
The initial step in developing the optimised phenyl block bases used in 
subsequent DMRG iterations. Dashed lines indicate the formation of reduced 
density matrices.
}
\label{diagram1}
\end{figure}

We embellish the basic approach described above with some further 
optimisations that serve to accelerate the convergence of energies with 
$m_A$ and $m_B$. Fig.\ \ref{diagram1} shows the construction of the initial 
phenyl bases: the end phenyl unit basis (for the first system block) is 
derived from the exact diagonalisation of biphenyl with open boundary 
conditions; the initial middle phenyl unit basis is derived from exactly 
diagonalising biphenyl with periodic boundary conditions. However, before 
using the middle block basis in DMRG iterations it pays to perform an 
improvement step using the trimer ($N = 3$). Namely, we diagonalise the 
timer using a relatively small number $m_A^{(0)}$ of states in the system 
block, and a large number of states $m_B^{(0)}$ in the middle block. A 
reduced density matrix is then formed for the middle unit (by performing a 
partial trace over the system and environment block Hilbert spaces on an 
appropriate projection operator) and diagonalised to give the basis used 
for middle units in the subsequent DMRG iterations.

\begin{figure}[htbp]
\centerline{\epsfxsize=7.5cm \epsfbox{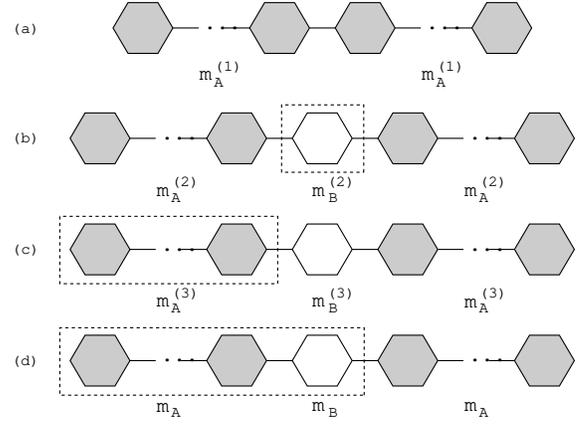}}
\caption{
The steps involved in a DMRG iteration: (a) Two system blocks are combined 
in order to treat systems with even $N$; (b) The middle phenyl unit basis 
is reoptimised in the context of this system size; (c) The system block is 
reoptimised for this superblock; (d) The final energy estimates are made, 
and the system block is augmented to produce the system block used in the 
next DMRG iteration. Dashed lines indicate the formation of reduced density 
matrices.
}
\label{diagram2}
\end{figure}

Fig.\ \ref{diagram2} shows a DMRG iterative step in the infinite lattice 
algorithm.
\begin{enumerate}

\item
First, a system block and its reflection are combined without a middle unit 
in order to treat systems with an even number of repeat units. In this 
calculation $m_A^{(1)}$ states are retained in the blocks. 

\item
Next, a middle unit is inserted in the superblock. However, we perform two 
further basis optimisations to improve the middle block and system block 
bases in the context of this system size before obtaining our final energy 
estimates.

\item
In the first step we improve the middle block basis (obtained from the 
trimer calculation depicted in Fig.\ \ref{diagram1}) by again retaining a 
relatively small number $m_A^{(2)}$ states in the system and environment 
blocks and a relatively large number $m_B^{(2)}$ states in the middle 
block, with $m_B^{(2)} \leq m_B^{(0)}$. A reduced density matrix is again 
formed for the middle block and diagonalised to give an improved middle 
block basis for this system size.

\item
A similar optimisation is then carried out for the system block, this time 
keeping a relatively small number of states $m_B^{(3)}$ in the middle unit 
and a relatively large number of states $m_A^{(3)}$ in the system and 
environment blocks. A reduced density matrix is formed for the system block 
by tracing out over the middle unit and environment block spaces.

\item
Finally, the energy estimates are obtained for this system size by 
retaining $m_A < m_A^{(3)}$ states from the reoptimised system block basis 
and $m_B < m_B^{(2)}$ states from the optimised middle block basis. It is 
also at this stage that the system block and middle unit are augmented to 
produce the system block used in the next iteration.

\end{enumerate}

Before discussing our results, we assess the convergence of the DMRG 
procedure with $m_A$ and $m_B$. We first consider the noninteracting limit 
($U = 0$) where comparisons can be made with exact results. Table 
\ref{conv0} shows the energies obtained for a number of states in various 
symmetry channels for $N = 11$ phenyl rings using $m_A = 280$ and $m_B = 
468$. In this example $m_A^{(0)} = 112$ and $m_B^{(0)} = 606$ and the extra 
middle unit and system block optimisations ((b) and (c) in Fig.\ 
\ref{diagram2}) are not used. The 
accuracy varies with the state studied but even the most inaccurate state 
is resolved to within 0.02 eV. These results are a number of orders of 
magnitude more accurate than a previous effort to apply the DMRG to the 
oligo-phenylene geometry \cite{ramasesha}.

In the interacting case of interest the convergence of energies is 
monitored as a function of both $m_A$ and $m_B$. We first consider biphenyl 
($N = 2$) in order to get an idea of the efficiency of the phenyl bases. In 
Table \ref{conv1} we list the energies of the two lowest $^1A_g^+$ states 
as functions of $m_A^{(1)}$, the number of system block states retained in 
step (a) of the procedure depicted in Fig.\ \ref{diagram2}. The indication 
is that even around 100 states around 3\% of the total basis gives a 
reasonably accurate truncated basis for the phenyl block. We next consider the 
initial trimer step ($N = 3$) depicted in Fig.\ \ref{diagram1}. In Table 
\ref{conv2} we list results for the $2^1A_g^+$ transition energy as a 
function of $m_A^{(0)}$ and $m_B^{(0)}$. The energy is converged to within 
around 0.01 eV. We have found that, once this initial reoptimisation of the 
middle phenyl block has been done for the trimer system, the parameter that 
controls the convergence (the parameter to which the energies are most 
sensitive) is the usual DMRG truncation parameter $m_A$, the number of 
states retained in the system and environment blocks. A typical example is 
given in Table \ref{conv3} where the $2^1A_g^+$ energy state is listed for 
various $m_A$ and $m_B$. The value of $m_A^{(0)}$ is set at 140 throughout, 
the value of $m_B^{(0)}$ is varied with $m_B$ as shown, and the extra 
middle unit and system block optimisations (steps (b) and (c) from Fig.\ \ref{diagram2}) are not used. The 
results indicate that the $2^1A_g^+$ energy is resolved to within around 
0.02 eV or better.

In production calculations we use all the embellishments depicted in Fig.\ 
\ref{diagram2}. A typical example of the parameters used is given by: 
$m_A^{(0)} = 168$, $m_B^{(0)} = 1000$, $m_A^{(1)} = 340$, $m_A^{(2)} = 
140$, $m_B^{(2)} = 740$, $m_A^{(3)} = 290$, $m_B^{(3)} = 300$, $m_A = 180$ 
and $m_B = 520$. We have not carried out a comprehensive study of the 
effect of the embellishments (b) and (c) but, from running a few sets of 
parameters for the same target states, and from observing the change in 
energy obtained from using a reduced, reoptimised basis for either the 
system block or the middle phenyl unit, the indication is that they allow 
us to work with effectively larger values of $m_A$ and $m_B$. That is, 
using the embellished procedure with the above parameter set is probably 
equivalent to the simpler scheme (without steps (b) and (c)) with $m_A 
\approx 290$ and $m_B \approx 740$. We are fairly confident that, in the 
calculations in this paper, the lowest lying eigenstates in a given 
symmetry channel are resolved to within around 0.02 eV or better \cite{footnote00}.

We now turn to a discussion of our results. We first embark on an 
investigation of the four characteristic absorption peaks of 
poly(phenylene) oligomers \cite{oligomers,zojer} and thin film polymers 
\cite{oligomers,absorption,lane}. Since the excitations of biphenyl ($N = 
2$) are relatively well understood, a knowledge of their evolution with 
chain length enables us to identify the key excitations in poly-phenyls. As 
the oligophenyls possess $D_{2h}$ symmetry, the dipole-active states are 
polarised either along the long axis ($^1B_{1u}^-$) or short axis 
($^1B_{2u}^-$). Fig.\ \ref{energies1} shows the $N$-dependence of 
transition energies of some key states, along with experimental results for 
biphenyl ($N = 2$) (see \cite{bursill98}), oligomers ($N = 3,\ldots,6$ 
\cite{oligomers} and $N = 6$ \cite{zojer}) and polymer thin films ($N = 
\infty$) \cite{oligomers,absorption,lane}.
We note that the particle-hole dipole-forbidden state ($1^1B_{2u}^+$) lies 
below the dipole-active $1^1B_{1u}^-$ state in biphenyl. However, the 
$1^1B_{2u}^+$ state very weakly hybridises, and thus its energy is almost 
independent of chain length, converging to 4.4 eV. This energy agrees very 
well with the very weak second absorption peak at 4.4--4.5 eV 
\cite{absorption,lane} in polymer thin films.
Adding weight to this interpretation is the observation of a weak but well 
defined 4.40 eV absorption peak in a highly textured film of sexiphenyl ($N 
= 6$), orientated perpendicular to the substrate \cite{zojer,footnote0}, as 
well as the weak, perpendicularly polarised absorption peak detected in 
orientated PFO film in the region 4.2--4.8 eV \cite{miller}. In contrast, 
the $1^1B_{1u}^-$ state strongly delocalises. As can be seen from Fig.\ 
\ref{energies1}, our DMRG results for the $1^1B_{1u}^-$ in the $N = 
3,\ldots,6$ systems practically coincide with oligomer data 
\cite{oligomers,zojer}. For large $N$ the $1^1B_{1u}^-$ energy approaches 
3.73 eV in reasonable agreement with the experimental peak observed at 
3.63--3.68 eV \cite{oligomers,lane}, or 3.2 eV \cite{absorption}. Together 
these results indicate that the first (strong) and second (weak) absorption 
peaks in phenyl-based systems are the $1^1B_{1u}^-$ and $1^1B_{2u}^+$ 
states, respectively.

The third and fourth absorption peaks are polarised normal ($^1B_{2u}^-$) 
and parallel ($^1B_{1u}^-$) to the long axis, respectively. In thin films 
they lie at 5.2--5.3 eV \cite{absorption,lane} and 5.7--6.0 eV 
\cite{absorption,lane} respectively. The $^1B_{1u}^-$ state is a localised 
intra-phenyl (Frenkel) excitation which lies at 6.16 eV in biphenyl 
\cite{bursill98}. To track the position of this state as a function of $N$, 
we perform calculations targeting a number of $^1B_{1u}^-$ states together 
with the ground state $1^1A_g^+$ and calculate the dipole moments
$\left\langle 1^1A_g^+ | \hat{\mu} | j^1B_{1u}^- \right\rangle$, where 
$\hat{\mu} \equiv \sum_i x_i n_i$ (with $x_i$ the $x$ coordinate of the 
$i$th atom) is the dipole operator along the long molecular axis. A typical 
set of dipole moments, for $N = 8$, is given in Table \ref{dipoles}. In 
general, it is found that there are two $^1B_{1u}^-$ states that are 
strongly dipole connected to the ground state. The first is the 
$1^1B_{1u}^-$, i.e., the first absorption peak. The second is the localised 
intra-phenyl excitation, i.e., the fourth absorption peak. (In the $N = 8$ 
case this is the $5^1B_{1u}^-$ state.) The $N$-dependence of this high-lying, strongly dipole connected $^1B_{1u}^-$ state is plotted in Fig.\ 
\ref{energies1}.

A conspicuous failure of the P-P-P model is its prediction for the 
remaining state, namely the lowest lying $^1B_{2u}^-$ state. The exact 
solution for biphenyl places this state at 6.66 eV, whereas experimentally 
it is at ca.\ 5.85 (and below the $2^1B_{1u}^-$ state). As shown in Fig.\ 
\ref{energies1} its calculated energy is 5.9 eV in the long chain limit, 
0.6--0.7 eV higher than the experimental value. 
Recent work has shown that the predicted energy of this state is improved 
if dielectric screening is introduced into the Ohno interaction 
\cite{castleton00}.

Fig.\ \ref{energies1} also shows the lowest lying triplet ($1^3B_{1u}^+$) 
and the dipole-forbidden $2^1A_g^+$ state. Electroabsorption studies place 
the $2^1A_g^+$ state at around 4.6 eV \cite{lane,footnote1}, around 0.5 eV 
below the extrapolated P-P-P result of 5.1 eV. (A more recent study 
\cite{absorption} places the $2^1A_g^+$ substantially lower). This 
discrepancy might possibly be explained in terms of the characteristic red 
shifts generally observed for certain excited states when going from well 
isolated chains to polymers in the solid state. Typical estimates for this 
polarisation or interchain screening shift are ca.\ 0.3 eV for the 
$^1B_{1u}^-$ state and ca.\ 0.6 eV for the $2^1A_g^+$ \cite{yaron}. As can 
be seen from Table \ref{dipoles} the $2^1A_g^+$ state has a large dipole 
moment with the $1^1B_{1u}^-$ state, and unlike the case for polyenes, it 
is not pre-dominantly a pair of bound magnons, but a particle-hole 
excitation. (It is usually labelled the $m^1A_g^+$ state.) This particle-hole excitation is either a `p'-wave exciton, or the edge of the unbound 
particle-hole continuum.

To investigate this further, and the nonlinear optical properties of 
poly(phenylene) in general, we consider dipole moments between various 
$^1A_g^+$ states and the $1^1B_{1u}^-$ state, as well as between the 
$2^1A_g^+$ state and various $^1B_{1u}^-$ states. The $N = 8$ values, 
listed in Table \ref{dipoles}, are representative of the general situation. 
We note that, in addition to the $1^1A_g^+$ and $2^1A_g^+$, another, higher 
lying state, which we denote the $k^1A_g^+$, also has an appreciable dipole 
moment with the $1^1B_{1u}^-$. (For the $N = 8$ case $k = 5$.) We also 
observed a pattern in the
$\left\langle 2^1A_g^+ | \hat{\mu} | j^1B_{1u}^- \right\rangle$ values. 
Namely, the $j=1$ state has a strong dipole moment with the $m^1A_g^+$, as 
does the higher lying $^1B_{1u}^-$ absorption peak state (localised intra-phenyl exciton). In addition, there is another state, lying higher still, 
that has the largest dipole moment with the $m^1A_g^+$. We adopt the usual 
convention of denoting this state as the $n^1B_{1u}^-$. (In the $N = 8$ 
case $n = 7$.)

\begin{figure}[htbp]
\centerline{\epsfxsize=7.5cm \epsfbox{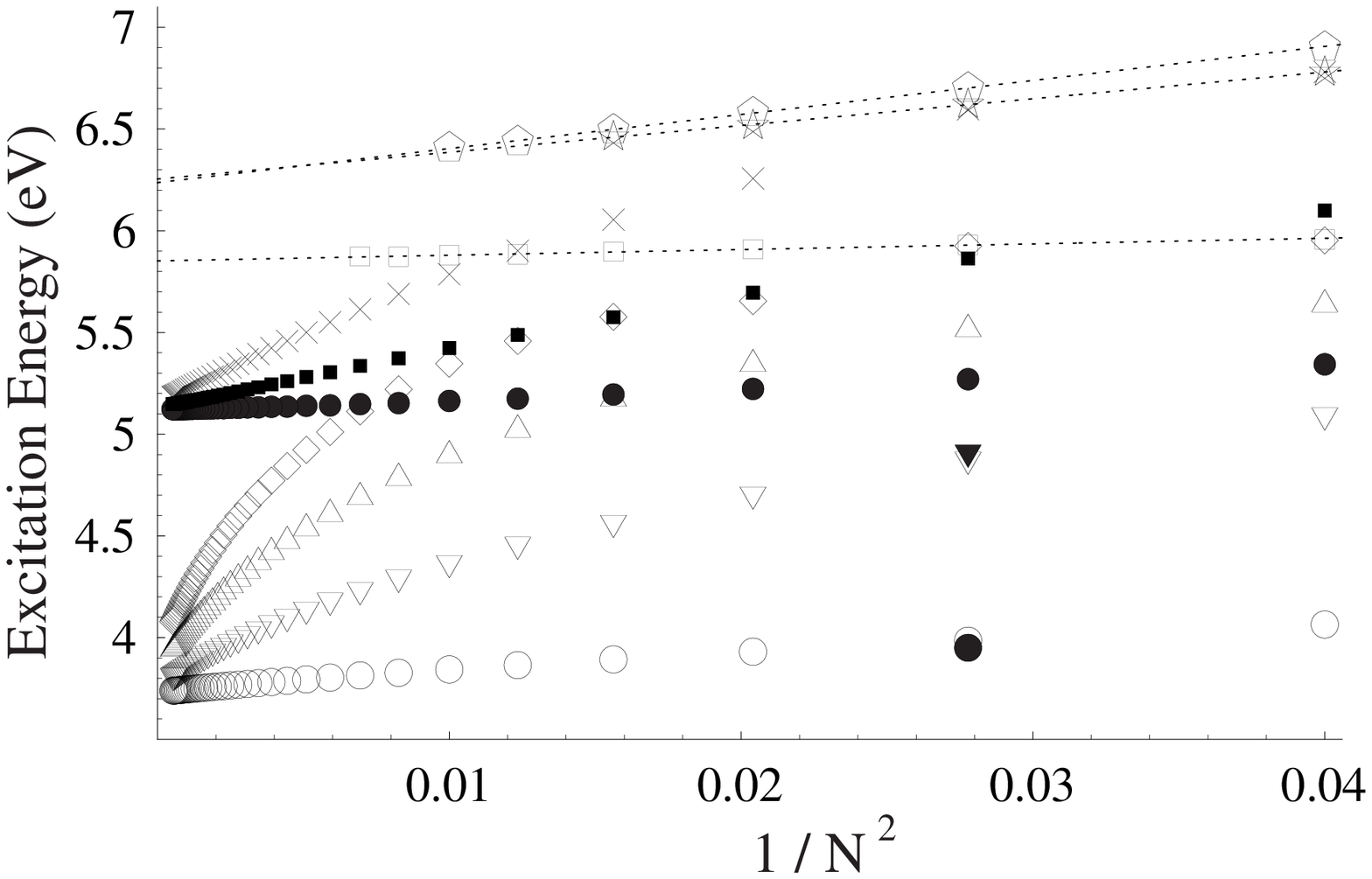}}
\caption{
The transition energies of a number of $^1A_g^+$ and $^1B_{1u}^-$ states as 
a function of $1 / N^2$, where $N$ is the number of repeat units: 
$1^1B_{1u}^-$ (large, open circles), $2^1B_{1u}^-$ (open, down triangles), 
$3^1B_{1u}^-$ (up triangles), $4^1B_{1u}^-$ (diamonds), the high lying 
localised intra-phenyl $^1B_{1u}^-$ excitation (open squares), $n^1B_{1u}^-
$ (pentagons), $m=2^1A_g^+$ (small, solid circles), $3^1A_g^+$ (small, 
solid squares), $4^1A_g^+$ ($\times$) and $k^1A_g^+$ (stars). The large, 
solid circle and solid down triangle show the position of the first and 
second long-axis polarised absoprtion peaks respectively for sexiphenyl ($N 
= 6$) \protect\cite{zojer,footnote0}. The dotted lines are to guide the 
eye.
}
\label{energies2}
\end{figure}

In Fig.\ \ref{energies2} we plot a number of $^1A_g^+$ and $^1B_{1u}^-$ 
state transition energies as functions of $1 / N^2$ \cite{footnote11}. We see that there are 
a number of states in the $^1B_{1u}^-$ sector which seem to converge to the 
same energy as the $1^1B_{1u}^-$ in the bulk limit. This is the band of 
`s'-wave charge-transfer excitons. Similarly, the $m^1A_g^+$ (i.e., the 
$2^1A_g^+$) has a number of momentum branches converging to the same 
energy, which is the band of `p'-wave charge-transfer excitons. Above this 
lies the localised intra-phenyl $^1B_{1u}^-$ state \cite{footnote2}. Higher 
in energy still are the $k^1A_g^+$ and the $n^1B_{1u}^-$ which appear to 
converge to the same energy in the $N = \infty$ limit. The strong dipole 
moments from the $m^1A_g^+$ to the $n^1B_{1u}^-$ and the $1^1B_{1u}^-$ to 
the $k^1A_g^+$, and the close proximity in energy of the $k^1A_g^+$ and the 
$n^1B_{1u}^-$ states indicate that these mark the onset of the continuum of 
unbound particle-hole excitations. Lying below this continuum are the `s'- 
and `p'-wave charge-transfer excitons and the Frenkel exciton.
This hypothesis could be checked further by calculating the average 
particle-hole separation for the various states to see if the $n^1B_{1u}^-$ 
is the first unbound state in the $^1B_{1u}^-$ sector. The convergence of 
the $k^1A_g^+$ and the $n^1B_{1u}^-$ energies to ca.\ 6.25 eV as $N 
\rightarrow \infty$ would imply a very large binding energy (ca.\ 2.5 eV) 
for the $^1B_{1u}^-$ exciton. However, band states are generally expected 
to be strongly effected by solid state screening (a red shift of 1.5 eV has 
been estimated for polyacetylene \cite{yaron}). Taking such a shift into 
account would bring the $n^1B_{1u}^-$ and hence the exciton binding energy 
much closer to the results implied by electroabsorption experiments 
\cite{absorption}.

In conclusion, we used a $\pi$-electron model to investigate the excited 
states of poly(phenylene), a model light emitting polymer. The transition 
energies were calculated using a new DMRG method. In order to accomodate 
the large repeat unit in the DMRG calculation, a truncated local Hilbert 
space was constructed for the phenyl unit. Using an optimal (or `natural') 
basis affords us convergence in the excitation energies to within a small 
fraction of an eV. For the most part, the calculated transition energies 
agree well with the experimental results for biphenyl. By monitoring the 
evolution of the energies with chain length and comparing them to the 
experimental absorption peaks of oligomers and thin film polymers, we can 
assign the four characteristic absorption peaks of phenyl-based light 
emitting polymers. In ascending order they are the $1^1B_{1u}^-$, 
$1^1B_{2u}^+$, $1^1B_{2u}^-$ and a high lying, localised intra-phenyl 
(Frenkel exciton) $1^1B_{1u}^-$ state. A failure of the P-P-P model is the 
incorrect prediction that the $1^1B_{2u}^-$ state lies above the Frenkel 
exciton \cite{footnote3}. We have also investigated the finite-size scaling 
of the nonlinear optical states in the longitudinally polarised one- 
($^1B_{1u}^-$) and the two- ($^1A_g^+$) photon sectors and found evidence 
for two $^1B_{1u}^-$ and one $^1A_g^+$ excitonic bands lying below the 
continuum.


Computations were performed at the New South Wales Center for Parallel 
Computing, The Australian Center for Advanced Computing and Communications 
and The Australian Partnership for Advanced Computation. Dr Bursill was 
supported by The Australian Research Council and The J. G. Russell 
Foundation.

\begin{table}[htbp]
\caption{
Comparison of DMRG and exact results for the $N = 11$ phenyl oligomer in 
the noninteracting ($U = 0$) limit. In the DMRG calculations $m_A = 280$ 
and $m_B = 468$.
}
\begin{tabular}{l|llll}
& $E(1^1A_g^+)$ & $E(1^1B_{1u}^-)$ & $E(2^1A_g^+)$ & $E(1^1B_{2u}^-)$ \\
\hline
DMRG  & $-231.0210234$ & $-228.4828$ & $-228.331$ & $-227.195$ \\
EXACT & $-231.0210256$ & $-228.4832$ & $-228.328$ & $-227.213$ \\
\end{tabular}
\label{conv0}
\end{table}

\begin{table}[htbp]
\caption{
Convergence of the lowest two states in the $^1A_g^+$ sector with 
$m_A^{(1)}$ for biphenyl ($N = 2$).
}
\begin{tabular}{l|ll}
\hline
$m_A^{(1)}$ & $E(1^1A_g^+)$ & $E(2^1A_g^+)$ \\
\hline
34    & $-63.0095953$ & $-56.7041530$ \\
92    & $-63.0125556$ & $-56.7137143$ \\
140   & $-63.0126802$ & $-56.7140886$ \\
188   & $-63.0126995$ & $-56.7141368$ \\
260   & $-63.0127066$ & $-56.7141589$ \\
322   & $-63.0127078$ & $-56.7141631$ \\
exact & $-63.0127086$ & $-56.7141653$ \\
\hline
\end{tabular}
\label{conv1}
\end{table}

\begin{table}[htbp]
\caption{
Convergence of the two-photon ($2^1A_g^+$) state transition energy with the 
DMRG parameters $m_A^{(0)}$ and $m_B^{(0)}$ for the trimer ($N = 3$).
}
\begin{tabular}{l|lll}
\hline
$m_B^{(0)}$ & $m_A^{(0)} = 34$ & $m_A^{(0)} = 92$ & $m_A^{(0)} = 140$ \\
\hline
420 & 5.7626 & 5.7498 & 5.7486 \\
574 & 5.7578 & 5.7449 & 5.7437 \\
692 & 5.7562 & 5.7433 & 5.7421 \\
952 & 5.7544 & 5.7415 & 5.7405 \\
\hline
\end{tabular}
\label{conv2}
\end{table}

\begin{table}[htbp]
\caption{
Convergence of the two-photon ($2^1A_g^+$) state transition energy with the 
DMRG parameters $m_A$ and $m_B$ for the $N = 11$ system.
}
\begin{tabular}{ll|llll}
\hline
$m_B$ & $m_B^{(0)}$ & $m_A = 34$ & $m_A = 85$ & $m_A = 130$ & $m_A = 165$ 
\\
\hline
34  & 140 & 5.38712 & 5.35049 & 5.36911 & 5.36862 \\
92  & 204 & 5.23364 & 5.17443 & 5.19057 & 5.18949 \\
140 & 322 & 5.21740 &   ---   & 5.16678 & 5.16533 \\ 
202 & 420 & 5.20744 &   ---   & 5.15356 & 5.15170 \\
268 & 574 & 5.20605 & 5.17076 & 5.15089 & 5.14903 \\
346 & 692 & 5.20490 & 5.16566 & 5.14948 & 5.14757 \\
418 & 848 & 5.20441 & 5.16545 & 5.14901 & 5.14715 \\
468 & 952 &   ---   & 5.16525 &   ---   & 5.14696 \\
\hline
\end{tabular}
\label{conv3}
\end{table}

\begin{table}[htbp]
\caption{
Dipole moments connecting various $^1A_g^+$ and $^1B_{1u}^-$ states for the 
$N = 8$ system.
}
\begin{tabular}{l|lll}
$j$ &
$\left\langle 1^1A_g^+ | \hat{\mu} | j^1B_{1u}^- \right\rangle$ &
$\left\langle j^1A_g^+ | \hat{\mu} | 1^1B_{1u}^- \right\rangle$ &
$\left\langle 2^1A_g^+ | \hat{\mu} | j^1B_{1u}^- \right\rangle$ \\
\hline
1 & 2.85 & 2.85 & 2.64 \\
2 & 0.68 & 2.64 & 0.48 \\
3 & 0.19 & 0.31 & 0.06 \\
4 & 0.11 & 0.14 & 0.02 \\
5 & 2.52 & 1.17 & 1.57 \\
6 & 1.03 &  --- & 1.31 \\
7 & 0.62 &  --- & 5.06 \\
8 & 0.48 &  --- & 0.04 \\
\end{tabular}
\label{dipoles}
\end{table}

\newpage

\begin{references}
\bibitem{footnote00}
Although extensive convergence tests have not been carried out for the even $N$ cases, in the noninteracting limit ($U = 0$) agreement with exact results is as good as that obtained for the case of odd $N$ (see Table \protect\ref{conv0}). Further, in the interacting case, energies are found to be smooth functions of $N$, so we are reasonably confident that our results for even $N$ are comparable in accuracy to those for odd $N$.
\bibitem{footnote0}
Although the $1^1B_{2u}^+$ state is particle-hole dipole forbidden in the 
P-P-P model, it is (weakly) observable in biphenyl (and presumably larger 
systems) due the fact that particle-hole symmetry is actually broken in 
real systems. Another possible interpretation of the second absorption peak 
is that it is due to the $2^1B_{1u}^-$ (see \protect\cite{chandross1}). That 
is, although the $2^1B_{1u}^-$ and $1^1B_{1u}^-$ coincide in the $N = 
\infty$ limit (see Fig.\ \protect\ref{energies2}), for systems of around $N 
= 8$ phenyl rings, the $2^1B_{1u}^-$ has an appreciable dipole moment with 
the $1^1A_g^+$ (see Table \ref{dipoles}) and has an energy of around 4.4 
eV. Although it is argued that this scenario is ruled out by polydispersity 
\protect\cite{weibel}, this alternative interpretation is possible if we 
assume that the conjugation length distribution is sharply peaked around $N 
= 8$ rings in thin films (i.e.\ unlike the $1^1B_{2u}^+$, the $2^1B_{1u}^-
$ has strong $N$-dependence). Interestingly, beyond the first maximum 
centered at 3.95 eV, in addition to the short-axis polarised peak at 4.4 
eV, the long-axis polarised absorption in \protect\cite{zojer} shows a peak 
at 4.91 eV that agrees well with the P-P-P $2^1B_{1u}^-$ result of 4.88 eV 
for $N = 6$ (see Fig.\ \protect\ref{energies2}).
\bibitem{footnote1}
It is usually assumed that the $2^1A_g^+$ state is a spin-density wave,
(as is the case for the dimerised Hubbard model in the strongly correlated regime) and hence does not contribute to the
nonlinear optical spectra. In this case, however, the $2^1A_g^+$ has a
large dipole moment with the $1^1B_{1u}^-$, and should thus be the
$m^1A_g^+$ state visible in electroabsortion.
\bibitem{footnote2}
We have seen some evidence of a band of states at this energy but it will 
be necessary to target a number of states in this energy range explicitly 
in order to clearly see higher momentum branches of the localised intra-phenyl $^1B_{1u}^-$ exciton.
\bibitem{footnote3}
Although the positions of these states can be inverted by reducing the 
electron correlation \protect\cite{chandross1} (the parameters in 
\protect\cite{chandross1} were determined within CI theory), or by the use 
of more phenomenological, semi-analytic approaches and a number of fitting 
parameters \protect\cite{rice}, our philosophy here has been to develop a 
numerical scheme that can first determine the predictions of P-P-P theory 
(without adjustable parameters or uncontrolled approximations), before 
embarking on the task of attempting to improve it.
\end{references}
\end{document}